\documentclass[11pt]{revtex4}
\usepackage{hyperref}
\usepackage{amsmath}

\usepackage{amsfonts}

\begin{document}
\title{\bf \Large    Reducible gauge theories in very special
relativity}

 \author { Sudhaker Upadhyay }
 \email{ sudhakerupadhyay@gmail.com;  sudhaker@iitk.ac.in}
\affiliation { Department of Physics, Indian Institute of Technology Kanpur, Kanpur 208016, India}

\begin{abstract} 
  In this paper we analyze the tensor field  (reducible gauge) theories in the context of very special relativity (VSR). Particularly, we study the VSR gauge symmetry as well as VSR BRST symmetry of
   Kalb-Ramond and Abelian 3-form fields involving a fixed null vector. We observe that
    the  Kalb-Ramond and Abelian 3-form fields and corresponding ghosts get masses in VSR framework. 
   The effective action in VSR-type axial gauge is greatly simplified compared with the VSR-type Lorenz gauge. Further, we quantize these models
   using a Batalin-Vilkovisy (BV) formulation in VSR.
  
\end{abstract}

\maketitle
 \section{Introduction}

Special relativity (SR)  postulates that the
laws of physics share many of the symmetries of Maxwell's equations and is valid at the largest energies attainable today
 \cite{pi}.  The maximal symmetry
group of Maxwell's equations is the   conformal
group $SU(2, 4)$.  However, the existence of particles with mass
 constrains spacetime symmetry to be no greater than the Lorentz group along with spacetime translations (i.e. Poincar\'e group).  The Poincar\'e 
group is proposed as the symmetry of nature by SR principles. 
 The possible violation of the
  Lorentz symmetry has received much attention  with new experimental
and theoretical challenges. For instance, 
many theories of quantum gravity predict breaking of some symmetry groups 
 \cite{alf}.    Experiments and astrophysical observations   set precise limits upon the parameters  illustrating these violations.
 The
spontaneous symmetry breaking of the Lorentz symmetry is assumed
  in extension of  the minimal standard model Refs. \cite{col,f}. 
 On the other hand,  the nondynamical tensor fields are
introduced to determine the preferred directions that break the
Lorentz symmetry. Some of such investigations   are discussed by the Myers-Pospelov model \cite{rc} together with QED in a constant axial vector background \cite{aa}. 
The
  Lorentz-invariant
theories  could emerge as effective
theories from a more fundamental scheme,  which is invariant under VSR groups but not invariant under the full
Poincar\'e group; this is  addressed in Ref. \cite{co}.

VSR is the set of subgroups of the Poincar\'e group which preserve the constancy of the velocity of light.
In this framework, it is   proposed that the laws of physics
are   invariant under the subgroups of Lorentz group SO(1,3) (with six parameters) rather than full Lorentz group    \cite{co,co1}.
Two most interesting
  subgroups of Lorentz group, the  four parameter   $SIM(2)$  group,
and the three parameter  $HOM(2)$ group have the property of rescaling a fixed null vector.
The remarkable property of these subgroups of the Lorentz
group is that  when they are supplemented
with $T, P$, or $CP$  the whole Lorentz group will be recovered. 
 VSR has been studied as regards  several aspects. For instance, it has been generalized for the inclusion of supersymmetry \cite{co2, vo}. It
admits the generation of a neutrino mass without lepton
number  and sterile neutrinos violations \cite{co1}. 
Further, it has been discussed in the case of
  curved spaces \cite{7,8},  noncommutativity
\cite{11}, the cosmological constant \cite{12}, dark matter
\cite{13}, cosmology \cite{14}, Abelian gauge fields \cite{15}, Born--Infeld electrodynamics \cite{16}, and non-Abelian gauge fields \cite{17}. Indeed it is not surprising that in spite of a considerable volume of research
on VSR, the higher form gauge theories in VSR  are still unstudied. A basic motivation of this paper is to
bridge this gap.

Higher-form gauge theories are generalizations of electromagnetism in which the
vector potential  a one-form  is replaced by exterior forms of higher degree.
Higher-form gauge theories are  an important ingredients in supergravity and superstring theory.
They are also   important  for the other branches of physics  \cite{green, pol, sud}.
For instance, the low energy excitations in string theories contain states described by antisymmetric tensor fields \cite{h,i}.  The antisymmetric tensor fields help to describe the various  
supergravity models.  The  Abelian rank-2
tensor field gets relevance for the classical string theories \cite{a}, for the theory of
vortex motion in an irrotational, incompressible fluid \cite{b,c}, for the 
dual formulation of the Abelian Higgs model \cite{d,e}.
They are also an ingredient of   supergravity multiplets 
\cite{g} and play a role in anomaly cancellation of certain superstring theories \cite{degu}.
The 3-form gauge fields are also important for supergravity theories.
For instant, $N=1$ supergravity theory in $d=11$ dimensions includes a massless 3-form gauge 
field. The study of solutions of such  supergravity theory shows that there are two-branes that indeed
do provide sources for 3-form fields \cite{st}.

In this paper we study the Kalb-Ramond (2-form) and Abelian rank-3 tensor (3-form) gauge theories in VSR.
Specifically,  we derive  the 
gauge-invariant action for such theories  in VSR. We notice that such an action is not invariant 
under the usual gauge transformation. However, it is invariant under the modified gauge 
transformation written in terms of the wiggle operator.  The equations of motion for a Kalb-Ramond field is derived by which one eventually
gets a mass in VSR. A guage theory cannot be quantized without 
choosing an appropriate gauge. Therefore,  we  choose a VSR-type Lorenz
gauge to quantize the theory. This gauge is incorporated in theory by adding a
corresponding  gauge-fixing term to the gauge-invariant action. 
To make the theory physically equivalent, the gauge-fixing term induce  a ghost term in the path 
integral. A remarkable feature of such study is that the ghost fields and ghost of ghost fields also get mass in 
VSR. Since all these fields   acquire a common mass, it cannot be used as an alternative for the Higgs mechanism.
Further we compute the BRST symmetry for Kalb-Ramond theory in VSR.
To quantize the theory a VSR-type axial  gauge is also chosen
which has a simpler form than the VSR-type  Lorenz  gauge. 
We   also quantize  the theory 
utilizing BV formulation where we derive the extended quantum action of the model
satisfying the quantum master equation.
Further we study an Abelian 3-form gauge theory in VSR.
The 3-form gauge field together with various ghost fields also acquires mass in the VSR framework. We further perform the BRST quantization of such model also in VSR.
We also shed light on Abelian 3-form gauge theory in a BV formulation with a
similar outcome as in case of the 2-form gauge theory. 

This presentation of the paper is as follows.
First we discuss the BRST quantization of an  Abelian 2-form gauge theory in 
VSR in section II. In this  section the  BV formulation prospects are also studied.
In section III, we analyze Abelian 3-form gauge theory in 
VSR. The BRST quantization and BV formulation are also studied in this section.
Finally, we conclude the results with a  remarks on future future in the last section.
 \section{Abelian 2-form fields in VSR}
 The Maxwell theory is modified in VSR; the same must happen to the
 Abelian rank-2 tensor (Kalb-Ramond) field theory.
To see this, we start with  
  the field-strength tensor in VSR for Kalb-Ramond tensor field $B_{\mu\nu}$ in VSR involving a fixed null vector $n_\mu$ as  
 \begin{eqnarray}
 F_{\mu\nu\rho}&=&\partial_\mu B_{\nu\rho}+\partial_\nu B_{\rho\mu}+\partial_\rho B_{\mu\nu}
+\frac{1}{2}m^2\left[n_\mu \frac{1}{(n\cdot\partial)^2} n^\alpha(\partial_\nu B_{\rho\alpha} +\partial_\rho B_{\nu\alpha}) \right.\nonumber\\
 &+&\left. n_\nu \frac{1}{(n\cdot\partial)^2} n^\alpha(\partial_\rho B_{\mu\alpha} +\partial_\mu B_{\rho\alpha})+n_\rho \frac{1}{(n\cdot\partial)^2} n^\alpha(\partial_\mu B_{\nu\alpha} +\partial_\nu B_{\mu\alpha})\right].\label{fi}
 \end{eqnarray}
 The null vector  $n^\mu$  transforms multiplicatively under a VSR transformation
so that  the terms containing ratios having  $n^\mu$ are invariant.
 This field-strength tensor  is not invariant under the standard gauge transformation 
 $ \delta B_{\mu\nu}= \partial_\mu \zeta_\nu - \partial_\nu\zeta_\mu$, where  $\zeta_{\mu}(x)$ is a vector  
parameter. 
Rather, this remains invariant under the following modified (VSR-type) gauge transformation: 
 \begin{eqnarray}
 \delta B_{\mu\nu}&=&\tilde\partial_\mu \zeta_\nu -\tilde\partial_\nu\zeta_\mu,\nonumber\\
 &=&\partial_\mu\zeta_\nu -\partial_\nu\zeta_\mu -\frac{1}{2}\frac{m^2}{n\cdot\partial}n_\mu\zeta_\nu +
 \frac{1}{2}\frac{m^2}{n\cdot\partial}n_\nu\zeta_\mu,
 \end{eqnarray}
 where $\tilde\partial_\mu=\partial_\mu  -\frac{1}{2}\frac{m^2}{n\cdot \partial}n_\mu$ is known as the wiggle operator. To have the usual mass dimension for the wiggle operator, a
constant $m$ has to be introduced which fixes the scale of
VSR effects.
  
The gauge-invariant action in VSR  describing the massive Kalb-Ramond tensor field is given by
 \begin{eqnarray}
 S^{(2)}_0=\frac{1}{12}\int d^4x\ \tilde F_{\mu\nu\rho}\tilde F^{\mu\nu\rho},\label{cl}
 \end{eqnarray}
 where the wiggle field-strength tensor has the following form:
 \begin{eqnarray}
 \tilde F_{\mu\nu\rho}&=&\tilde \partial_\mu B_{\nu\rho}+\tilde\partial_\nu B_{\rho\mu}+\tilde\partial_\rho B_{\mu\nu},\nonumber\\
 &=&\partial_\mu B_{\nu\rho}+\partial_\nu B_{\rho\mu}+\partial_\rho B_{\mu\nu}
 -\frac{1}{2}\frac{m^2}{n\cdot \partial}n_\mu B_{\nu\rho}-\frac{1}{2}\frac{m^2}{n\cdot \partial}n_\nu B_{\rho\mu}-\frac{1}{2}\frac{m^2}{n\cdot \partial}n_\rho B_{\mu\nu},\nonumber\\
 &=&F_{\mu\nu\rho}  -\frac{1}{2} m^2 \left(n_\mu\frac{1}{(n\cdot\partial)^2}n^\alpha F_{\nu\rho\alpha}+n_\nu\frac{1}{(n\cdot\partial)^2}n^\alpha F_{ \rho\mu\alpha}+n_\rho\frac{1}{(n\cdot\partial)^2}n^\alpha F_{\mu\nu\alpha} \right).
 \end{eqnarray}
It is evident from the above relation that $\tilde F_{\mu\nu\rho}$ does not coincide with  $  F_{\mu\nu\rho}$ given in (\ref{fi}).
 
 The equations of motion (EOM) for the Kalb-Ramond field is calculated as
 \begin{eqnarray}
 \tilde\partial_{\mu}\tilde F^{\mu\nu\rho}=0.
 \end{eqnarray}
 For the VSR-type Lorenz gauge $\tilde \partial_\mu B^{\mu\nu}=0$, the EOM reduces to
 \begin{eqnarray}
 [\square -m^2 ]B^{\nu\rho}=0,
 \end{eqnarray}
which remarkably implies that the field $B_{\mu\nu}$ has mass $m$. The non-local terms
are dealt with the following relation \cite{ale}:
\begin{eqnarray}
\frac{1}{n\cdot\partial}=\frac{1}{\partial_t +\partial_z}=\int  dt_+,
\end{eqnarray}
where $t_+=\frac{t+z}{2}$.
Here we observe that our results are in agreement with \cite{15}. 
Next we will study the covariant quantization of Abelian 2-form gauge theory in VSR.
 \subsection{Different gauges}
 In order to quantize a gauge theory we must add gauge fixing
term and the corresponding Faddeev-Popov term to the invariant action. Doing so, the gauge fixing term breaks the local gauge symmetry 
and thus  the divergence of the functional integral disappears. However, the ghost term improves
the integration measure to provide correct predictions for gauge invariant observables. Therefore, for  
so-called BRST quantization, it is necessary to introduce the following 
ghost and auxiliary fields for reducible 2-from gauge theory: anticommuting vector fields $\rho_{\mu}$ and $\bar\rho_{\mu}$, 
a commuting vector field $\beta_{\mu}$, anticommuting scalar fields $\chi$ and $\bar\chi$, 
and commuting scalar fields $\sigma, \varphi,$ and $ \bar\sigma $.  
The gauge fixing and ghost action for antisymmetric rank 2 tensor field in VSR-type Lorenz gauge is given by
\begin{eqnarray}
S_{gf+gh}^{(2)L}&=&\int d^4x\left[ i\bar\rho_\nu \tilde\partial_\mu(\tilde\partial^\mu\rho^\nu -
\tilde\partial^\nu\rho^\mu )-\bar\sigma\tilde\partial_\mu\tilde\partial^\mu\sigma +\beta_\nu(\tilde\partial_\mu B^{
\mu\nu} +\lambda_1\beta^\nu -\partial^\nu\varphi)\right.\nonumber\\ 
&-&\left. i\bar\chi(\tilde\partial_\mu\rho^\mu +\lambda_2 \chi) -i\bar\rho^\mu \tilde \partial_\mu \chi \right],\nonumber\\
&=&\int d^4x\left[i\bar\rho_\nu \left(\partial_\mu\partial^\mu \rho^\nu -\partial_\mu\partial^\nu
\rho^\mu -m^2\rho^\nu +\frac{1}{2}	\frac{m^2}{n\cdot \partial}n^\nu\partial\cdot\rho
+ \frac{1}{2}	\frac{m^2}{n\cdot \partial} \partial^\nu n\cdot\rho\right.\right.\nonumber\\
& -&\left.\left. \frac{1}{4}\frac{m^2}{(n\cdot\partial)^2}n^\nu n\cdot\rho\right) -\bar{\sigma}
(\partial_\mu\partial^\mu -m^2)\sigma +\beta_\nu\partial_\mu B^{
\mu\nu} -\frac{1}{2}m^2\beta_\nu\frac{1}{n\cdot\partial}n_\mu B^{\mu\nu}+\lambda_1\beta_\nu\beta^\nu \right.\nonumber\\
&-& \left. \beta_\nu\partial^\nu\varphi -i\bar\chi \partial_\mu\rho^\mu +\frac{i}{2}m^2\bar\chi\frac{1}{n\cdot\partial}n_\mu \rho^\mu-i\lambda_2\bar\chi\chi -i\bar\rho^\mu\partial_\mu\chi-\frac{i}{2}\frac{m^2}{n\cdot\partial}\bar\rho^\mu n_\mu\chi\right], \label{gfix}
\end{eqnarray}
where $\lambda_1$ and $\lambda_2$ are gauge parameters.
It is evident from the above expression that the ghost fields and ghost of ghost fields
have   mass $m$ in  VSR. Since all the fields   acquire a common mass, it cannot be used as a replacement for the Higgs mechanism.
The ghost propagator and ghost of ghost propagator are computed, respectively, as
\begin{eqnarray}
&&D_{\mu\nu}^{gh}(k) =-\frac{1}{k^2+m^2}\left[g_{\mu\nu}+\frac{k_\mu k_\nu}{m^2}\right],\nonumber\\
&&D^{ggh}(p) =-\frac{1}{p^2+m^2}.
\end{eqnarray}
It can be seen that   the propagators and
vertices have the same large momentum behavior as in
Lorentz-invariant  theories. So the 2-form gauge theory in VSR is renormalizable.

The expression (\ref{gfix}) can further be written in terms of BRST variation $\delta_b$
of gauge-fixing fermion $\psi^L$ as follows:
\begin{eqnarray}
S_{gf+gh}^{(2)L}&=&\delta_b\int d^4x\ \psi^L,\nonumber\\
&=&\delta_b\int d^4x\left[-i\bar\rho_\nu(\tilde\partial_\mu B^{\mu\nu}+\lambda_1\beta^\nu-\tilde\partial^\nu\varphi)
-i\bar\sigma(\tilde\partial_\mu\rho^\mu +\lambda_2\chi)\right],\nonumber\\
&=&\delta_b\int d^4x\left[-i\bar\rho_\nu  \partial_\mu B^{\mu\nu}+\frac{i}{2}m^2\bar\rho_\nu\frac{1}{n\cdot \partial}n_\mu B^{\mu\nu}-i\lambda_1\bar\rho_\nu\beta^\nu +i\bar\rho_\nu\partial^\nu\varphi 
 \right.\nonumber\\
&-&\left.  \frac{i}{2}m^2\bar\rho_\nu\frac{1}{n\cdot \partial}n^\nu\varphi -i\bar\sigma\left(\partial_\mu\rho^\mu -\frac{1}{2}\frac{m^2}{n\cdot \partial}n_\mu \rho^\mu +\lambda_2\chi\right)\right],\label{gff}
\end{eqnarray}
where the BRST transformation of the fields is given by
\begin{eqnarray}
\delta_b B_{\mu\nu} &=& -\left(\partial_\mu\rho_\nu -\partial_\nu\rho_\mu -\frac{1}{2}\frac{m^2}{n\cdot\partial}n_\mu\rho_\nu +\frac{1}{2}\frac{m^2}{n\cdot\partial}n_\nu\rho_\mu\right)\Lambda, \nonumber\\
\delta_b\rho_\mu &=& -i\left( \partial_\mu\sigma - \frac{1}{2}\frac{m^2}{n\cdot\partial}n_\mu\sigma\right)\Lambda,    \ \ \ \ \delta_b\sigma 
= 0, \nonumber\\
\delta_b\bar\rho_\mu &=&i\beta_\mu \Lambda,    \ \ \ \ 
\delta_b\beta_\mu = 0,\  \ \ \
\delta_b\bar\sigma =-\bar\chi\Lambda, \nonumber\\
\delta_b\varphi &=& -\chi\Lambda, \ \ \ \ \ 
\delta_b\bar\chi =0, \ \ \ \ \delta_b\chi =0.\label{sym}
\end{eqnarray}
Here $\Lambda$ is an infinitesimal Grassmann parameter.

The gauge fixing and ghost action for the antisymmetric rank-2 tensor field in VSR-type axial gauge
 (i.e. $\eta_\mu B^{\mu\nu}=0$) is given by
 \begin{eqnarray}
S_{gf+gh}^{(2)A}&=&\int d^4x\left[ i\bar\rho_\nu \eta_\mu(\tilde\partial^\mu\rho^\nu -
\tilde\partial^\nu\rho^\mu )-\bar\sigma\eta_\mu\tilde\partial^\mu\sigma +\beta_\nu(\eta_\mu B^{
\mu\nu} +\lambda_1\beta^\nu -\eta^\nu\varphi)\right.\nonumber\\ 
&-&\left. i\bar\chi(\eta_\mu\rho^\mu +\lambda_2 \chi) -i\bar\rho^\mu \eta_\mu \chi \right],\nonumber\\
&=&\int d^4x\left[ i\bar\rho_\nu \eta_\mu\left(\partial^\mu\rho^\nu -
\partial^\mu\rho^\nu  -\frac{1}{2}\frac{m^2}{n\cdot \partial}n^\mu\rho^\nu +\frac{1}{2}\frac{m^2}{n\cdot \partial}n^\mu\rho^\nu\right)-\bar\sigma\eta_\mu\partial^\mu\sigma \right.\nonumber\\ 
&+&\left.\frac{1}{2}m^2 
\bar\sigma \frac{1}{n\cdot\partial}\eta\cdot n\sigma +\beta_\nu(\eta_\mu B^{
\mu\nu} +\lambda_1\beta^\nu -\eta^\nu\varphi)
- i\bar\chi(\eta_\mu\rho^\mu +\lambda_2 \chi) -i\bar\rho^\mu \eta_\mu \chi \right].
\end{eqnarray}
In terms of a gauge fixing fermion it can further be written as
\begin{eqnarray}
S_{gf+gh}^{(2)A}&=&\delta_b\int d^4x\left[-i\bar\rho_\nu(\eta_\mu B^{\mu\nu}+\lambda_1\beta^\nu-\eta^\nu\varphi)
-i\bar\sigma(\eta_\mu\rho^\mu +\lambda_2\chi)\right].
\end{eqnarray}
We see here that the gauge-fixed action in VSR-type axial  gauge has simpler form than the
Lorenz gauge. Next we discuss BV formulation of this model.
\subsection{Batalin-Vilkovisky formulation}
To analyze the BV formulation for Abelian rank-2 antisymmetric
tensor field theory  in VSR, we  first define the 
generating functional in the VSR-type Lorenz gauge in field/antifield formulation by introducing 
an antifield  corresponding to each field of the theory  with opposite statistics, thus:
\begin{eqnarray}
Z_{2-form}^L &=&\int\left[dBd\rho d\bar{\rho}d\sigma d\bar{\sigma}d\varphi d\chi d\bar{\chi}d
\beta\right]\exp\left[i\int d^4x\left\{\frac{1}{12}F_{\mu\nu\lambda}F^{\mu\nu\lambda}\right.\right. \nonumber\\
&-& \left.\left.  B^{
\mu\nu\star}\left(\partial_\mu\rho_\nu-\partial_\nu\rho_\mu -\frac{1}{2}\frac{m^2}{n\cdot\partial}n_\mu\rho^\nu +\frac{1}{2}\frac{m^2}{n\cdot\partial}n_\nu\rho^\mu\right)
\right.\right. \nonumber\\
&-& \left.\left. i\rho^{\mu\star}\left(\partial_\mu\sigma - \frac{1}{2}\frac{m^2}{n\cdot\partial}n_\mu\sigma\right) +i{\bar{\rho}}^{\nu\star}\beta_\nu-\bar{
\sigma}^\star\bar\chi-\varphi^\star\chi\right\}\right].
\end{eqnarray}
These antifields (starred fields) are identified  with the help of
the gauge fixed fermion given in (\ref{gff}) as 
\begin{eqnarray}
\psi^L
&=&  -i\bar\rho_\nu  \partial_\mu B^{\mu\nu}+\frac{i}{2}m^2\bar\rho_\nu\frac{1}{n\cdot \partial}n_\mu B^{\mu\nu}-i\lambda_1\bar\rho_\nu\beta^\nu +i\bar\rho_\nu\partial^\nu\varphi -\frac{i}{2}m^2\bar\rho_\nu\frac{1}{n\cdot \partial}n^\nu\varphi
 \nonumber\\
&-&  i\bar\sigma\left(\partial_\mu\rho^\mu -\frac{1}{2}\frac{m^2}{n\cdot \partial}n_\mu \rho^\mu +\lambda_2\chi\right).
\end{eqnarray}
These identifications are
 \begin{eqnarray}
B^{\mu\nu\star }&=&\frac{\delta \psi^L}{\delta B_{\mu\nu}}=i\partial^\mu\bar\rho^\nu+
\frac{i}{2}m^2\bar\rho^\nu\frac{1 }{n\cdot\partial}n^\mu,  \nonumber\\ 
\bar\rho^{\nu\star}&=&\frac{\delta \psi^L}{\delta \bar \rho_\nu}=-i\left(\partial_\mu B^{\mu\nu}-\frac{1}{2}m^2\frac{1}{n\cdot\partial}n_\mu B^{\mu\nu}+
\lambda_1\beta^\nu -\partial^\nu\varphi +\frac{1}{2}m^2\frac{1}{n\cdot\partial}n^\nu\varphi \right),\nonumber\\
\rho^{\mu\star }&=&\frac{\delta \psi^L}{\delta \rho_{\mu}}=i\partial^\mu\bar\sigma +\frac{i}{2}\bar\sigma\frac{m^2}{n\cdot\partial}n^\mu,  \ \ \ \
\bar\sigma^\star =\frac{\delta \psi^L}{\delta\bar\sigma}=-i\left(\partial_\mu\rho^\mu -\frac{1}{2}\frac{m^2}{n\cdot \partial}n_\mu \rho^\mu +\lambda_2\chi\right),\nonumber\\
\sigma^\star &=&\frac{\delta \psi^L}{\delta \sigma}=0, \ \ \ \ \ \ \ \chi^\star =\frac{\delta \psi^L}{\delta\chi}=-i\lambda_2 \bar \sigma,\nonumber\\
\varphi^\star &=&\frac{\delta \psi^L}{\delta\varphi}=-i
\partial_\mu\bar\rho^\mu -\frac{i}{2}m^2\bar\rho_\nu\frac{1}{n\cdot\partial}n^\nu, \ \ \bar\chi^\star =\frac{\delta \psi^L}{
\delta\bar\chi}=0.
\end{eqnarray}
This  can further be expressed  in a compact form as
\begin{equation}
Z^L_{2-form} = \int {\cal D}\phi\ e^{iW_{2-form}(\phi,\phi^\star)},
\end{equation}
where $W_{2-form}(\phi,\phi^\star)$ is an extended quantum action for the Abelian 2-form gauge theory in 
the VSR-type Lorenz gauge  written in terms of generic field $\phi$ and antifield $\phi^\star$. It is well-known that the value of generating functional $Z^L_{2-form}$ does not depend on the choice of gauge-fixing fermion.
This extended quantum action, $W_{2-form}(\phi,\phi^\star)$, is the solution of  certain rich mathematical
relation, which is called the quantum master equation, given by
\begin{equation}
\Delta e^{iW_{2-form}[\phi, \phi^\star ]} =0,\ \
 \Delta\equiv (-1)^{\epsilon}\frac{\partial_l}{
\partial\phi}\frac{\partial_l}{\partial\phi^\star } .
\end{equation}
Corresponding to different choices of gauge condition, there will be many
possible solutions of the quantum master equation.
The ghost number and statistics of $\phi^\star$ 
are
\begin{eqnarray}
\mbox{gh} [\phi^\star]=-\mbox{gh} [\phi]-1,\  \ \ \epsilon(\phi^\star)= \epsilon(\phi)+1\ (\mbox{mod} 
\ 2).
\end{eqnarray}
 The quantum action can be extended up to the one-loop order correction
   as
   \begin{equation}
  W_{2-form}[\phi, \phi^\star ]=S_0[\phi] +S^{(2)L}_{gf+gh} [\phi, \phi^\star ] +\hbar M_1[\phi, \phi^\star ],
   \end{equation}
where $S_0 +S^{(2)L}_{gf+gh} $ is the complete action   given in Eqs. (\ref{cl}) and (\ref{gff}) and $M_1$ appears from nontrivial measure factors.    
     
   The behavior of $ W_{2-form}$  for BRST 
   transformations can be given by 
   \begin{equation}
  \delta_b
 W_{2-form} =i\hbar\Delta  W_{2-form}.
   \end{equation}
   For (non-anomalous) gauge theory up to first-order correction $M_1$ 
  the solution does not depend on antifields. 
   In this situation, the BRST transformations of the complete 
   action $S_0 +S^{(2)L}_{gf+gh}$ and $M_1$    are given by
   \begin{equation}
   \delta_b (S_0 +S^{(2)L}_{gf+gh})=0, \ \ \delta_b M_1 =i\Delta (S_0 +S^{(2)L}_{gf+gh}).
   \end{equation}
This result can further be generalized up to higher order of perturbation.  
  
In the next section we will study the   case of an Abelian rank-3 tensor field theory in VSR.
\section{Abelian 3-form fields in VSR}
The Abelian 3-form gauge field is important for supergravity theory in higher spacetime dimensions.
So it is important to study such a gauge field in VSR.
Let us start by writing the  the field-strength for the Abelian 3-form gauge theory  in 
  arbitrary $d$ dimensions
for VSR as
\begin{eqnarray}
H_{\mu\nu\eta\chi}&=&\partial_\mu B_{\nu\eta\chi} -\partial_\nu B_{\eta\chi\mu}+
\partial_{\eta}B_{\chi\mu\nu}-\partial_\chi B_{\mu\nu\eta}+\frac{1}{2}m^2 \left[n_\mu 
\frac{1}{(n\cdot\partial)^2}n^\alpha (\partial_\nu B_{\eta\chi\alpha
}-\partial_\eta B_{\chi\alpha\nu}+\partial_\chi B_{\alpha\nu\eta})\right.\nonumber\\
&-&\left. n_\nu 
\frac{1}{(n\cdot\partial)^2}n^\alpha (\partial_\eta B_{ \chi\mu\alpha
}-\partial_\chi B_{\mu\alpha\eta}+\partial_\mu B_{\alpha \eta\chi})+n_\eta
\frac{1}{(n\cdot\partial)^2}n^\alpha (\partial_\chi B_{\mu\nu\alpha
}-\partial_\mu B_{\nu\alpha\chi}+\partial_\nu B_{\alpha\chi\mu})\right.\nonumber\\
&-&\left. n_\chi
\frac{1}{(n\cdot\partial)^2}n^\alpha (\partial_\mu B_{ \nu\eta\alpha
}-\partial_\nu B_{\eta\alpha\mu}+\partial_\eta B_{\alpha\mu\nu})
\right].
\end{eqnarray}
It is straightforward to check that this field-strength is not invariant under the
standard gauge transformation, $\delta B_{\mu\nu\eta}=\partial_\mu\lambda_{\nu\eta}+\partial_\nu\lambda_{\eta\mu}+\partial_\eta
\lambda_{\mu\nu}$.
Rather, this  is invariant under the following modified (VSR-type) gauge
transformation: 
\begin{eqnarray}
\delta B_{\mu\nu\eta}=\partial_\mu\lambda_{\nu\eta}+\partial_\nu\lambda_{\eta\mu}+\partial_\eta
\lambda_{\mu\nu}-\frac{1}{2}\frac{m^2}{n\cdot\partial}n_\mu\lambda_{\nu\eta}-\frac{1}{2}\frac{m^2}{n\cdot\partial}n_\nu\lambda_{\eta\mu}-\frac{1}{2}\frac{m^2}{n\cdot\partial}n_\eta\lambda_{\mu\nu},
\label{gau}
\end{eqnarray}
where  $\lambda_{\mu\nu}$ is  a tensor parameter of transformation.

To describe a massive 3-form field we define the VSR-type gauge invariant action in $d$ dimensions as follows:
\begin{eqnarray}
S_0= \kappa\int d^dx \ \tilde H_{\mu\nu\eta\chi}\tilde H^{\mu\nu\eta\chi},\label{ac}
\end{eqnarray}
where $\kappa$ is some fixed constant and the wiggle field strength for the 3-form gauge field is given by
\begin{eqnarray}
\tilde H_{\mu\nu\eta\chi}&=&\tilde \partial_\mu B_{\nu\eta\chi} -\tilde \partial_\nu B_{\eta\chi\mu}+
\tilde \partial_{\eta}B_{\chi\mu\nu}-\tilde \partial_\chi B_{\mu\nu\eta},\nonumber\\
&=&H_{\mu\nu\eta\chi}+\frac{1}{2}m^2\left[n_\mu\frac{1}{(n\cdot\partial)^2}n^\alpha H_{\nu\eta\chi\alpha}
- n_\nu\frac{1}{(n\cdot\partial)^2}n^\alpha H_{ \eta\chi\mu\alpha}\right.\nonumber\\
&+&\left. n_\eta\frac{1}{(n\cdot\partial)^2}n^\alpha H_{ \chi\mu\nu\alpha}-n_\chi\frac{1}{(n\cdot\partial)^2}n^\alpha H_{\mu\nu\eta\alpha} \right].
\end{eqnarray}
From the above expression it is evident that the wiggle field strength does not coincide
with the field strength $H_{\mu\nu\eta\chi}$.
Now, the EOM for the 3-form gauge field is calculated by 
\begin{equation}
\tilde{\partial}_\mu\tilde{H}^{\mu\nu\eta\chi} =0,
\end{equation}
which, in turn, for VSR-type Lorenz gauge (i.e. $\tilde{\partial}_\mu B^{\mu\nu\eta}$)   reduces to
\begin{equation}
(\square -m^2)B^{\nu\eta\chi}=0.
\end{equation}
This is a Klein-Gordon equation for a massive field.
This implies that the 3-form gauge field $B^{\nu\eta\chi}$ has mass $m$.

Since the action (\ref{ac}) respects the VSR-type gauge symmetry, for a perturbative formulation,
we need to break the local gauge invariance by adding a gauge-fixing term. When constructing
the effective action at higher orders, maintaining unitarity, one has to replace the
local guage symmetry by (global) BRST symmetry. To make the
gauge-fixing term BRST invariant, we need to add ghost terms to the effective action.
We, therefore, fix a VSR-type Lorenz gauge (i.e. $\tilde{\partial}_\mu B^{\mu\nu\eta}=0$).
Since it is a reducible gauge theory, 
we need some more fixing for other (ghost) fields.
So, this gauge-fixing condition is incorporated by adding the following 
gauge-fixed action together with the induced ghost term:
\begin{eqnarray}
S^L_{gf+gh} 
&=& \int d^dx \left[ \tilde\partial_\mu B^{\mu\nu\eta}B_{\nu\eta} +
\frac{1}{2}B_{\mu\nu}\bar B^{\mu\nu} 
+
(\tilde\partial_\mu \bar c_{\nu\eta} + \tilde\partial_\nu \bar c_{\eta\mu}
+ \tilde\partial_\eta \bar c_{\mu\nu})\tilde\partial ^\mu 
c^{\nu\eta}\right.\nonumber\\ 
&-&  \left.(\tilde\partial_\mu\bar \beta_\nu -\tilde\partial_\nu \bar\beta_\mu )\tilde\partial^\mu\beta^\nu -BB_2 -
   \frac{1}{2} B_1^2 +(\tilde\partial_\mu \bar c^{\mu\nu}+\tilde\partial^\nu \bar c_1)f_\nu  \right.
\nonumber\\
&-&\left.(\tilde\partial_\mu c^{\mu\nu}- \tilde{\partial}^\nu c_1)\bar F_\nu +\tilde\partial_\mu\bar c_2 \tilde\partial^\mu c_2 
 + \tilde\partial_\mu\beta^\mu B_2 +\tilde\partial_\mu \phi^\mu B_1 -
\tilde\partial_\mu\bar\beta^\mu B\right],
\end{eqnarray}
where antisymmetric ghost    and   antighost fields ($c_{\mu\nu}$ and $\bar c_{\mu\nu}$) are   
Grassmannian and  the vector field 
 $\phi_\mu$, antisymmetric auxiliary fields  $B_{\mu\nu}, \bar B_{\mu\nu}$  and  auxiliary fields  $B, 
B_1, B_2$  are bosonic in nature. The ghost of ghosts ($\beta_\mu$ and $\bar\beta_\mu$) are bosonic in nature. However, ghost of ghost of ghosts ($c_2$ and $\bar{c}_2$) are
fermionic in nature. The rest of the Grassmannian fields ($c_1, \bar c_1, f_\mu$ and $\bar F_\mu$) are  auxiliary fields.  It can easily be see here that 
the ghosts ($c_{\mu\nu}$ and $\bar c_{\mu\nu}$), ghost of ghosts ($\beta_\mu$ and $\bar\beta_\mu$) and ghost of ghost of ghosts ($c_2$ and $\bar{c}_2$) have mass $m$.

Expanding the wiggle operation, it reduces to
\begin{eqnarray}
S^L_{gf+gh} &=&  \int d^dx \left[  \partial_\mu B^{\mu\nu\eta}B_{\nu\eta} -\frac{1}{2}m^2B_{\nu\eta}\frac{1}{n\cdot\partial}n_\mu B^{\mu\nu\eta}+
\frac{1}{2}B_{\mu\nu}\bar B^{\mu\nu} 
+
( \partial_\mu \bar c_{\nu\eta} +  \partial_\nu \bar c_{\eta\mu}
+  \partial_\eta \bar c_{\mu\nu}) \partial ^\mu 
c^{\nu\eta}\right.\nonumber\\ 
&+&  \left.m^2\bar c_{\nu\eta}c^{\nu\eta} -m^2\partial_\nu\bar c_{\eta\mu}\frac{1}{n\cdot\partial}
n^\mu c^{\nu\eta} -\frac{m^2}{n\cdot\partial}n_\nu\bar c_{\eta\mu}\partial^\mu c^{\nu\eta}+
\frac{1}{2}\frac{m^4}{n\cdot\partial}n_\nu\bar{c}_{\eta\mu}\frac{1}{n\cdot\partial}n^\mu c^{\nu\eta}
\right.\nonumber\\ 
&-&  \left.( \partial_\mu\bar \beta_\nu -  \partial_\nu \bar\beta_\mu ) \partial^\mu\beta^\nu -m^2\bar \beta_\nu\beta^\nu -\frac{1}{2}m^2\partial_\nu\bar\beta_\mu\frac{1}{n\cdot\partial}n^\mu\beta^\nu 
-\frac{1}{2}\frac{m^2}{n\cdot\partial}n_\nu\bar\beta_\mu\partial^\mu\beta^\nu 
\right.
\nonumber\\
&+&\left.\frac{1}{4}\frac{m^4}{n\cdot\partial}n_\nu\bar\beta_\mu \frac{1}{n\cdot\partial}n^\mu\beta^\nu 
-BB_2 -
   \frac{1}{2} B_1^2 +(\partial_\mu \bar c^{\mu\nu})f_\nu -\frac{1}{2}\frac{m^2}{n\cdot\partial}n_\mu\bar c^{\mu\nu}f_\nu \right.
\nonumber\\
&+&\left. \partial^\nu \bar c_1 f_\nu - \frac{1}{2}\frac{m^2}{n\cdot\partial}n^\nu \bar c_1 f_\nu -( \partial_\mu c^{\mu\nu})\bar F_\nu  +\frac{1}{2}\frac{m^2}{n\cdot\partial}n_\mu c^{\mu\nu}\bar F_\nu
+\partial^\nu c_1 \bar F_\nu  
  \right.
\nonumber\\
&-&\left.   \frac{1}{2}\frac{m^2}{n\cdot\partial}n^\nu c_1 \bar F_\nu   -\bar c_2( \partial_\mu\partial^\mu -m^2)c_2  
  +  \partial_\mu\beta^\mu B_2-\frac{1}{2}\frac{m^2}{n\cdot\partial}n_\mu\beta^\mu B_2 \right.
\nonumber\\
&+&\left.   \partial_\mu \phi^\mu B_1-\frac{1}{2}\frac{m^2}{n\cdot\partial}n_\mu   \phi^\mu B_1-
 \partial_\mu\bar\beta^\mu B+ \frac{1}{2}\frac{m^2}{n\cdot\partial}n_\mu \bar\beta^\mu B\right].\label{lag}
\end{eqnarray}

The effective action together with (\ref{ac}) and (\ref{lag}) remains invariant under 
the following set of BRST transformations:
\begin{eqnarray}
\delta_b B_{\mu\nu\eta} &=& -\left(\partial_\mu c_{\nu\eta}+\partial_\nu c_{\eta\mu} +\partial_\eta c_{\mu\nu}-\frac{1}{2}\frac{m^2}{n\cdot\partial}n_\mu c_{\nu\eta}-\frac{1}{2}\frac{m^2}{n\cdot\partial}n_\nu c_{ \eta\mu}-\frac{1}{2}\frac{m^2}{n\cdot\partial}n_\eta c_{\mu\nu}\right) \Lambda,\nonumber\\
\delta_b c_{\mu\nu}  &=& \left(\partial_\mu\beta_\nu -\partial_\nu \beta_\mu-\frac{1}{2}\frac{m^2}{n\cdot\partial}n_\mu \beta_\nu +\frac{1}{2}\frac{m^2}{n\cdot\partial}n_\nu \beta_\mu\right) \Lambda,\
\delta_b\bar c_{\mu\nu}=B_{\mu\nu} \Lambda, \nonumber\\
\delta_b\bar B_{\mu\nu}  &=&-\left(\partial_\mu f_\nu -\partial_\nu f_\mu-\frac{1}{2}\frac{m^2}{n\cdot\partial}n_\mu f_\nu +\frac{1}{2}\frac{m^2}{n\cdot\partial}n_\nu f_\mu\right) \Lambda,   \ \ 
\delta_b\bar\beta_\mu = -\bar F_\mu \Lambda,\nonumber\\
\delta_b\beta_\mu &=&-\left(\partial_\mu  c_2 -\frac{1}{2}\frac{m^2}{n\cdot\partial}n_\mu c_2\right)\Lambda,  \ \ \
\delta_b\bar c_2 =B_2 \Lambda,\ \ \  \ \  \delta_b c_1=-B \Lambda,\nonumber\\
 \delta_b \phi_\mu 
 &=&-f_\mu \Lambda,\ \
\delta_b \bar c_1=  B_1  \Lambda,\ \ \
\delta_b {\cal M} =0,\ \ \ \
{\cal M}   \equiv   \{c_2, f_\mu, \bar F_\mu, B, B_1, B_2, B_{\mu\nu}\},\label{brst}
\end{eqnarray}
where $\Lambda$ is the fermionic transformation parameter.
The gauge fixing fermion is given by
\begin{eqnarray}
\psi_L &=&  -\partial_\mu\bar c_{\nu\eta}B^{\mu\nu\eta}+\frac{1}{2}\frac{m^2}{n\cdot\partial}
n_\mu \bar c_{\nu\eta} B^{\mu\nu\eta} -\frac{1}{2}\bar c_2 B 
+\frac{1}{2}c_1B_2
 -  \frac{1}{2}\bar c_1 B_1 - c^{\mu\nu}
\partial_\mu\bar \beta_\nu +c_1\partial_\mu \bar{\beta}^\mu  \nonumber\\
&+&\frac{1}{2}c^{\mu\nu}\frac{m^2}{n\cdot\partial}n_\mu\bar\beta_\nu - \partial_\mu \bar c_2 \beta^\mu +\frac{1}{2}\frac{m^2}{n\cdot\partial}n_\mu \bar c_2 \beta^\mu + \frac{1}{2}\bar c_{\mu\nu} \bar B^{\mu\nu}  +\bar c_1\partial_\mu \phi^\mu -\frac{1}{2}\bar{c}_1 \frac{m^2}{n\cdot\partial}n_\mu \phi^\mu. \label{psi}
\end{eqnarray}
This expression will play an important role in the next subsection 
to get identification for the antifields in VSR-type Lorenz gauge.
\subsection{Batalin-Vilkovisky formulation}
To describe 3-form gauge theory in BV formulation in VSR, we
introduce the antifields corresponding to  each field of the model with opposite 
statistics having non-vanishing 
BRST symmetry in the generating functional as follows:
 \begin{eqnarray}
Z_{3-form}^L &=&\int {\cal D}\phi \exp\left[i\int d^dx\left\{\frac{1}{24}F_{\mu\nu\eta\chi}F^{\mu\nu\eta\chi}
-
B_{\mu\nu\eta}^\star \left(
\partial^\mu c^{\nu \eta} + \partial^\nu c^{  \eta\mu} +
\partial^\eta c^{\mu\nu }\right. \right.\right. \nonumber\\
 & -& \left.\left.\left.  \frac{1}{2}\frac{m^2}{n\cdot\partial}n_\mu c_{\nu\eta}-\frac{1}{2}\frac{m^2}{n\cdot\partial}n_\nu c_{ \eta\mu}-\frac{1}{2}\frac{m^2}{n\cdot\partial}n_\eta c_{\mu\nu} \right)+
    {c}_{\mu\nu}^\star \left( \partial^\mu\beta^\nu  -\partial^\nu\beta^\mu \right. \right.\right. \nonumber\\
 & -& \left.\left.\left.  \frac{1}{2}\frac{m^2}{n\cdot\partial}n_\mu \beta_\nu +\frac{1}{2}\frac{m^2}{n\cdot\partial}n_\nu \beta_\mu
    \right)+\bar{c}_{\mu\nu}^\star 
 B^{\mu\nu}  -\bar B_{\mu\nu}^\star \left(\partial^\mu f^\nu   -\partial^\nu f^\mu  \right. \right.\right. \nonumber\\
 & -& \left.\left.\left. \frac{1}{2}\frac{m^2}{n\cdot\partial}n_\mu f_\nu +\frac{1}{2}\frac{m^2}{n\cdot\partial}n_\nu f_\mu\right) 
-\beta_\mu^\star \left(\partial^\mu c_2   -\frac{1}{2}\frac{m^2}{n\cdot\partial}n_\mu c_2  \right) -\bar \beta_\mu^\star \bar F ^\mu \right.\right. \nonumber\\
 &  +& \left.\left.
\bar c_2^\star   B_2  
 +\bar c_1^\star  B_1 -c_1^\star  B 
-\phi_\mu ^\star   f^\mu \right\}\right ].
\end{eqnarray}
These antifields (starred fields) are evaluated with the help of the gauge-fixing fermion (\ref{psi}) as follows
\begin{eqnarray}
&&B_{\mu\nu\eta}^\star=-\partial_\mu \bar c_{\nu\eta}+\frac{1}{2}\frac{m^2}{n\cdot\partial}
n_\mu \bar c_{\nu\eta} \ \
  c_{\mu\nu}^\star  = -\partial_\mu\bar \beta_\nu+\frac{1}{2} \frac{m^2}
  {n\cdot\partial}n_\mu\bar\beta_\nu,\ \
\bar  B_{\mu\nu}^\star = \frac{1}{2}\bar  c_{\mu\nu},\nonumber\\
&&\bar  c_{\mu\nu}^\star =\frac{1}{2}\bar  B_{\mu\nu} +
\partial^\eta B_{\mu\nu\eta}-\frac{1}{2}\frac{m^2}{n\cdot\partial}
n^\eta   B_{\eta\mu\nu},\ \
 \beta_{\mu}^\star =- \partial_\mu\bar  c_2+\frac{1}{2}\frac{m^2}{n\cdot\partial}n_\mu \bar c_2,
 \nonumber\\
&&\bar \beta_{\mu}^\star =-\partial_\mu c_1 +\partial^\nu c_{\nu\mu}-\frac{1}{2}c_{\nu\mu}\frac{m^2}
{n\cdot\partial}n^\nu,\ \ \bar  c_2^{\star}=-\frac{1}{2}B+\partial_\mu\beta^\mu -\frac{1}{2} \frac{m^2}
{n\cdot\partial}n_\mu\beta^\mu,\nonumber\\ 
&&
  c_1^{\star} =\frac{1}{2}B_2 +\partial_\mu \bar \beta^\mu,\ \
\bar  c_1^{\star} =-\frac{1}{2}B_1 +\partial_\mu\phi^\mu -\frac{1}{2}\bar{c}_1 \frac{m^2}{n\cdot\partial}n_\mu \phi^\mu,\nonumber\\
&&\phi_{\mu}^\star =-\partial_\mu \bar c_1 -\frac{1}{2}\frac{m^2}
{n\cdot\partial}n_\mu,\ \
 B^{\star} =-\frac{1}{2}\bar  c_2,\ \
  B_1^{\star} =-\frac{1}{2}\bar  c_1,\nonumber\\
&&B_2^{\star} = \frac{1}{2}c_1,\ \
 \{  B_{\mu\nu}^\star, \bar  F_{\mu}^\star, f_{\mu }^\star,   c_2^{\star}\}=0.
\end{eqnarray}
The generating functional can further  be written in compact form as
\begin{equation}
Z^L_{3-form} = \int {\cal D}\phi\ e^{iW_{3-form}(\phi,\phi^\star)},
\end{equation}
where $W_{3-form}(\phi,\phi^\star)$ is an extended quantum action for Abelian 3-form gauge theory in 
the VSR-type Lorentz gauge  written in terms of generic field $\phi$ and antifield $\phi^\star$.
This extended quantum action, $W_{3-form}(\phi,\phi^\star)$, is the solution of a certain rich mathematical
relation, which is called the  quantum master equation, given by
\begin{equation}
\Delta e^{iW_{3-form}[\phi, \phi^\star ]} =0,\ \
 \Delta\equiv (-1)^{\epsilon}\frac{\partial_l}{
\partial\phi}\frac{\partial_l}{\partial\phi^\star } .
\label{mq}
\end{equation}
Here $W_{3-form}[\phi, \phi^\star ]$ is the solution of quantum master equation.
For different gauge choices, it corresponds to different solutions of the 
quantum master equation.
From this  quantum master equation one can get a relation between different 
correlation functions. 
 \section{Conclusion}
In this paper we have analyzed the reducible gauge theories in VSR. To be more specific, we
have demonstrated the Kalb-Ramond field theory in VSR involving a fixed null vector. We have derived the
classical action for such theory in VSR. We have found that the action is not invariant 
under the standard gauge transformation for the Kalb-Ramond field. However, such action remains invariant under the modified (VSR-type) gauge 
transformation written in terms of
the wiggle operator.  We have derived the equations of motion for a Kalb-Ramond field, which eventually
turns out to be the Klein-Gordon equation for a massive field.
This ensures that the Kalb-Ramond field in VSR gets a mass.
gets a mass. Further, to quantize such a theory in VSR we have fixed a VSR-type Lorenz
gauge which breaks the local (VSR) gauge symmetry. The propagators have also been calculated.
This gauge-fixing term   induced a ghost term in the path integral.
Here we have observed that the ghost fields and ghost of ghost fields also get a common mass in VSR.
Therefore it cannot be an alternative to the Higgs mechanism. 
Further we demonstrate the BRST symmetry for Kalb-Ramond gauge theory in VSR.
To break the gauge symmetry an axial-type gauge has also been chosen
which has simpler form than the Lorenz type gauge. We have also quantized the theory 
utilizing BV formulation where we derive the extended quantum action of the model to the first 
order in perturbation
satisfying the quantum master equation.

Subsequently, we have considered an Abelian 3-form gauge theory (another reducible gauge theory) also in VSR which plays an important role in $11$ dimensional supergravity.
It has been shown that this model also respects a VSR-type gauge invariance rather than standard gauge symmetry.
We have found that the gauge fields together with all ghost fields get s common mass for s such theory in VSR. We have also analyzed the BRST quantization of 3-form gauge theory in VSR.
Further, we have studied the model in  BV formulation. 
Now it would be extremely interesting to evaluate the different identities following the BRST symmetry of the reducible gauge theories in VSR. This might be helpful in gaining a clear understanding of the theory in VSR. It will 
also be interesting to explore such 
results in further interesting models such as perturbative quantum gravity,
super-Yang-Mills theory and supersymmetric Chern-Simmons theory etc.

\end{document}